\documentclass[12pt]{article}
\usepackage{amsmath,amssymb,amsfonts, epsfig}
\usepackage{color}
\usepackage{bm}

\makeatletter

\@addtoreset{equation}{section}
\makeatother

\begin{document}
\title{
\begin{flushright}
\ \\*[-80pt] 
\begin{minipage}{0.2\linewidth}
\normalsize
KEK-TH-1780
\end{minipage}
\end{flushright}
{\Large \bf 
Quark mixing from $\Delta (6N^2)$ family symmetry 
\\*[20pt]}}
\author{
Hajime~Ishimori,$^{1}$ 
Stephen~F.~King,$^{2}$ 
Hiroshi~Okada,$^{3}$ 
{Morimitsu\,Tanimoto}$^{4}$ \\
\\*[20pt]
\centerline{
\begin{minipage}{\linewidth}
\begin{center}
$^1${\it \normalsize 
Institute of Particle and Nuclear Studies,\\
High Energy Accelerator Research Organization (KEK)\\
Tsukuba 305-0801, Japan} \\
$^2${\it \normalsize
School of Physics and Astronomy,
University of Southampton,\\
Southampton, SO17 1BJ, U.K.}\\
$^3${\it \normalsize
School of Physics, KIAS, Seoul 130-722, Korea}\\
$^4${\it \normalsize
Department of Physics, Niigata University, Niigata 950-2181, Japan}\\
\end{center}
\end{minipage}}
\\*[50pt]}
\vskip 2 cm
\date{\small
\centerline{ \bf Abstract}
\begin{minipage}{0.9\linewidth}
\medskip 
We consider a direct approach to quark mixing based on the discrete
family symmetry $\Delta (6N^2)$
in which the Cabibbo angle is determined
by a residual $Z_2\times Z_2$ subgroup to be $|V_{us}|=0.222521$,
for $N$ being a multiple of 7. 
We propose a particular model in which 
unequal smaller quark mixing angles and CP phases may occur without breaking the residual $Z_2\times Z_2$
symmetry. We perform a numerical analysis of the model for $N=14$, where 
small $Z_2\times Z_2$ breaking effects of order 3\% 
are allowed by model, allowing perfect agreement within the uncertainties of the 
experimentally determined best fit quark mixing values.
\end{minipage}
}

\begin{titlepage}
\maketitle
\thispagestyle{empty}
\end{titlepage}

\section{Introduction}

Non-Abelian discrete groups have been extensively used as family symmetries in the lepton sector,
in order to account for the large leptonic mixing angles~\cite{pdg}
(for reviews see e.g.~\cite{Altarelli:2010gt,Ishimori:2010au,King:2013eh,King:2014nza}.)
In the direct approach, a non-Abelian family symmetry in the lepton sector is assumed.
Following the determination of a Cabibbo-sized reactor angle, the 
only viable class appears to be $\Delta (6N^2)$ for large $N$ values 
\cite{Holthausen:2012wt,King:2013vna,Lavoura:2014kwa,Fonseca:2014koa}.
Then, such a symmetry is broken to 
 $Z_2 \times Z_2$ in the neutrino sector (the so called Klein symmetry) and  $Z_3$
 in the charged lepton sector, with the mixing angles determined from symmetry. 

An analogous approach based on $\Delta (6N^2)$ has also been considered in the quark sector
\cite{Araki:2013rkf,Ishimori:2014jwa}.
In the quark sector one may envisage a residual
$Z_n\times Z_m$ symmetry of the quark mass matrices,
where this is a subgroup of the $\Delta (6N^2)$ family symmetry.
However, in the quark sector, this approach is more 
challenging due to the small mixing angles. 
Nevertheless, earlier work showed that the Cabibbo angle could 
emerge from a residual $Z_2\times Z_2$ symmetry,
arising as a subgroup of the dihedral family symmetry 
$D_{7}$ \cite{Lam:2007qc,Blum:2007jz}, 
$D_{12}$ \cite{Kim:2010zub}, or $D_{14}$ \cite{Blum:2009nh,Blum:2007nt,Hagedorn:2012pg}. 
Then, more general analyses based on larger discrete family symmetry groups
was considered~\cite{Holthausen:2013vba,Araki:2013rkf}. 
Some authors have speculated that both the lepton mixing angles and the Cabibbo angle
may arise from some common discrete family symmetry group
\cite{Hagedorn:2012pg,Holthausen:2013vba}. 
Note that only the Cabibbo angle is determined, since the residual $Z_2\times Z_2$ symmetry
only fixes the upper $2\times 2$ block of the mixing matrix. 
The Cabibbo angle is predicted by $\theta_C=\pi n/N$ where 
$n$ and $N$ are integers relating to the family symmetry. 
A complementary approach to deriving the Cabibbo angle of $\theta_{C}\approx 1/4 $
at leading order was recently considered in an indirect model
based on a vacuum alignment $(1,4,2)$ without any residual symmetry
\cite{King:2013hoa}.

It is clear that the residual $Z_2\times Z_2$ symmetry is insufficient by itself
to determine all the small quark mixing angles. Moreover, it is not even sufficient to 
fully determine the structure of the CKM matrix, since the eigenvalues of $Z_2$ are $\pm 1$,
hence at least two eigenvalues of the $3\times 3$ generators
should be the same. In order to break the degeneracy, it is necessary to
consider concrete models. 
In a recent paper \cite{Ishimori:2014jwa}, a realistic model of quarks was proposed based on the discrete
family symmetry $\Delta (6N^2)$, where the residual symmetry for the quark sector was assumed 
to be $Z_2\times Z_2$ symmetry, corresponding to a $Z_2$ symmetry in each of the up and down sectors. 
However, a drawback of that model was that, the resulting structure of the CKM matrix
required $\theta_{23}= \theta_{13}$, in the $Z_2\times Z_2$ symmetry limit.
The purpose of the present paper is to consider an alternative direct model of quarks based on 
$\Delta (6N^2)$ in which an alternative $Z_2\times Z_2$ subgroup is preserved which allows $\theta_{23}\neq \theta_{13}$.
As in the previous model, the present model will provide a qualitative explanation for 
the smaller mixing angles, although their quantitative values must be  
fitted to experimental values, rather than being predicted.

This paper is organized as follows.
In section 2,  we discuss the $Z_n\times Z_m$ symmetry of the quark mass matrices
and the relation with the CKM matrix.
In section 3, we present  a brief review  of the group theory of the $\Delta (6N^2)$ series and identify suitable $Z_2\times Z_2$ subgroups which may be preserved in the quark sector, leading to a successful determination of the Cabibbo angle. In section 4, we present a model of quarks based on  $\Delta (6N^2)$. We construct the quark mass matrices and resulting CKM mixing  and derive the vacuum alignments that are required.
In section 5, we perform a full numerical analysis of the model for $N=14$
and show that all the quark masses, CKM mixing angles and the
 unitarity triangle are accommodated.
Section 6 is devoted to the summary.

\section{CKM matrix and $Z_n\times Z_m$ symmetry of quark mass matrices}
\label{2}
The quark mass matrices, $M_u$ and $M_d$,  are defined in a general RL basis by
\begin{align}
-{\cal L}=
\begin{pmatrix}
\overline{u} & \overline{c} & \overline{t}
\end{pmatrix}_R
M_u 
\begin{pmatrix}
u\\
c\\
t\\
\end{pmatrix}_L
 + 
 \begin{pmatrix}
\overline{d} & \overline{s} & \overline{b}
\end{pmatrix}_R
M_d
\begin{pmatrix}
d\\
s\\
b\\
\end{pmatrix}_L + H.c..
\end{align}
We write the mass matrices in the diagonal basis with hats, where,
\begin{equation}
M_u = V_u' {\hat M}_u V_u^\dagger  \ \ {\rm and} \ \
 M_d = V_d' {\hat M}_d V_d^\dagger .
\end{equation}
Hence,
\begin{equation}
M_u^\dagger M_u = V_u {\hat M}_u^\dagger {\hat M}_u V_u^\dagger  \ \ {\rm and} \ \
 M_d^\dagger M_d = V_d {\hat M}_d^\dagger {\hat M}_d V_d^\dagger .
\end{equation}
Thanks to $Z_n\times Z_m$ symmetry,  the quark mass matrices in the diagonal basis are invariant under ${\hat Q}$ and ${\hat A}$ transformations, 
\begin{equation}
{\hat Q}^\dagger  \left({\hat M_u}^\dagger  {\hat M_u} \right) {\hat Q} = {\hat M_u}^\dagger  {\hat M_u} \ \ {\rm and} \ \
{\hat A}^\dagger  \left({\hat M_d}^\dagger  {\hat M_d} \right) {\hat A} = {\hat M_d}^\dagger  {\hat M_d} ,
\end{equation}
where ${\hat Q}$ and ${\hat A}$ are elements of $Z_n$ and $Z_m$, respectively, given by
\begin{eqnarray}
{\hat Q}=\begin{pmatrix}
e^{2\pi i n_u/n}&0&0\\
0&e^{2\pi i n_c/n}&0\\
0&0&e^{2\pi i n_t/n}\\
\end{pmatrix},
\quad
{\hat A}=\begin{pmatrix}
e^{2\pi i m_d/m}&0&0\\
0&e^{2\pi i m_s/m}&0\\
0&0&e^{2\pi i m_b/m}\\
\end{pmatrix},
\end{eqnarray}
where $n_{u,c,t}$ and $m_{d,s,b}$ are integers. 
It then follows that in the original (non-diagonal) basis that the mass matrices are invariant under $Q$ and 
$A$ transformations,
\begin{equation}
Q^\dagger  \left(M_u^\dagger  M_u \right) Q = M_u^\dagger  M_u \ \ {\rm and} \ \
A^\dagger  \left( M_d^\dagger  M_d \right) A = M_d^\dagger  M_d ,
\end{equation}
where 
\begin{eqnarray}
Q=V_u {\hat Q}
V_u^\dagger,
\quad
A=V_d {\hat A}
V_d^\dagger .
\end{eqnarray}
In the non-diagonal basis they also satisfy $Q^n=A^m= e$. Since the CKM matrix is given by $V_u^\dagger V_d$, up to phase transformations, it can be determined
from the matrices which diagonalise $Q$ and $A$,
\begin{eqnarray}
Q=V_Q {\hat Q}
V_Q^\dagger,
\quad
A=V_A {\hat A}
V_A^\dagger ,
\end{eqnarray}
where we identify $V_u=V_Q$ and $V_d=V_A$.

\section{The group $\Delta(6N^2)$ and $Z_2$ symmetry}
\label{3}

Let us briefly review the discrete group $\Delta (6N^2)$
\cite{Ishimori:2010au}, which is  isomorphic 
to $(Z^c_N \times Z^d_N)\rtimes S_3$.
The group $S_3$ is isomorphic to $Z^a_3\rtimes Z^b_2$, 
where we denote the generators of $Z^a_3$ and $Z^b_2$ as $a$ and $b$ 
and we write the generators of $Z^c_N$ and $Z^d_N$ as $c$ and $d$. 
These generators  satisfy
\begin{eqnarray}
& & a^3=b^2=(ab)^2 =c^N = {d}^N =  e, 
\quad cd = dc, 
\nonumber \\
& &  aca^{-1}=c^{-1}d^{-1}, \quad ada^{-1}=c,
\nonumber \\
& &  bcb^{-1}=d^{-1}, \quad bdb^{-1}=c^{-1}.
\end{eqnarray}
Using them, all of $\Delta (6N^2)$ elements are written as 
\begin{eqnarray}
 & & g=a^{k}b^{\ell}c^{m}d^{n},
\end{eqnarray}
for $k=0,1,2$, $\ell=0,1$ and $m,n=0,1,2,\cdots ,N-1$.

For $N/3\not=$integer, irreducible representations are 
${\bf 1}_{0,1}$, $\bf 2$, ${\bf 3}_{1k}$, 
${\bf 3}_{2k}$, and ${\bf 6}_{[[k],[\ell]]}$. 
Tensor products relating to doublet and triplets are 
\begin{eqnarray}
\begin{split}
{\bf 3}_{1k}\times {\bf 3}_{1k'}
={\bf 3}_{1(k+k')}+{\bf 6}_{[[k],[-k']]},
\quad
{\bf 3}_{1k}\times {\bf 3}_{2k'}
={\bf 3}_{2(k+k')}+{\bf 6}_{[[k],[-k']]},
\\
{\bf 3}_{2k}\times {\bf 3}_{2k'}
={\bf 3}_{1(k+k')}+{\bf 6}_{[[k],[-k']]},
\quad
{\bf 3}_{1k}\times {\bf 2}
={\bf 3}_{1 k }+{\bf 3}_{2k},
\\
{\bf 3}_{2k}\times {\bf 2}
={\bf 3}_{1 k }+{\bf 3}_{2k},
\quad
{\bf 2}\times {\bf 2} 
={\bf 1}_{0}+{\bf 1}_{1}+{\bf 2}.
\end{split}
\end{eqnarray}
Some triplets and sextet are reducible, 
precisely ${\bf 3}_{10}={\bf 1}_0+{\bf 2}$, 
${\bf 3}_{20}={\bf 1}_1+{\bf 2}$, 
and ${\bf 6}_{[[-k],[k]]}={\bf 3}_{1k}+{\bf 3}_{2k}$. 
If their representations are explicitly given, they are 
$(x_1,x_2,x_3)_{{\bf 3}_{10}}=(x_1+x_2+x_3)_{{\bf 1}_0}
+(\omega x_1+x_2+\omega^2 x_3,\omega^2 x_1+x_2+\omega x_3)_{\bf 2}$, 
$(x_1,x_2,x_3)_{{\bf 3}_{20}}=(x_1+x_2+x_3)_{{\bf 1}_1}
+(\omega x_1+x_2+\omega^2 x_3,\omega^2 x_1+x_2+\omega x_3)_{\bf 2}$, 
and 
$(x_1,x_2,x_3,x_4,x_5,x_6)_{{\bf 6}_{[[-k],[k]]}}
=(x_1+x_6,x_2+x_5,x_3+x_4)_{{\bf 3}_{1k}}
+(-x_1+x_6,-x_2+x_5,-x_3+x_4)_{{\bf 3}_{2k}}$. 

In a particular matrix representation, the irreducible triplet generators are,
\begin{eqnarray}
a=\begin{pmatrix}
0&1&0\\
0&0&1\\
1&0&0
\end{pmatrix},
\quad
b=\pm\begin{pmatrix}
0&0&1\\
0&1&0\\
1&0&0
\end{pmatrix},
\quad
c=\begin{pmatrix}
\eta^k&0&0\\
0&\eta^{-k}&0\\
0&0&1
\end{pmatrix},
\quad
d=\begin{pmatrix}
1&0&0\\
0&\eta^k&0\\
0&0&\eta^{-k}
\end{pmatrix},
\end{eqnarray}
for the triplet ${\bf 3}_{1k}$ with plus sign 
and for ${\bf 3}_{2k}$ with minus sign 
where $\eta=e^{2\pi i/N}$.

Let us consider $Q=abc^{x}$ and $A=abc^{y}$, i.e.
\begin{align}
\label{qanda}
Q=
\begin{pmatrix}
0&\eta^{-kx}&0\\
\eta^{kx}&0&0\\
0&0&1
\end{pmatrix},
\quad
A=
\begin{pmatrix}
0&\eta^{-ly}&0\\
\eta^{ly}&0&0\\
0&0&1
\end{pmatrix},
\end{align}
for ${\bf 3}_{1k}$ to $Q$ and ${\bf 3}_{1l}$ to $A$. 
Because of the degeneracy of the two eigenvalues $+1$ for the above 
matrices,we generally have
\begin{align}
Q=
V_Q
\begin{pmatrix}
\pm1&0&0\\
0& \mp1&0\\
0&0&+1
\end{pmatrix}
V_Q^\dagger,
\quad
A=
V_A
\begin{pmatrix}
\pm 1&0&0\\
0& \mp1&0\\
0&0&+1
\end{pmatrix}
V_A^\dagger,
\end{align}
which corresponds to having a $+1$ eigenvalue in the (3,3) position and the other two eigenvalues $\pm 1$
being in all possible places, with the trace equal to $+1$. The position of  these eigenvalues is not fixed by symmetry arguments alone since they may be interchanged by further (1,2) unitary rotations,
with each choice being consistent with $Q,A$ in Eq. (\ref{qanda}).
A particular model will resolve the degeneracy. For example in the model in \cite{Ishimori:2014jwa},
the ordering chosen was,
\begin{align}
Q=
V_Q
\begin{pmatrix}
-1&0&0\\
0&1&0\\
0&0&1
\end{pmatrix}
V_Q^\dagger,
\quad
A=
V_A
\begin{pmatrix}
-1&0&0\\
0&1&0\\
0&0&1
\end{pmatrix}
V_A^\dagger,
\end{align}
This ordering was responsible for the unwanted prediction $\theta_{23}= \theta_{13}$, 
as discussed in 
\cite{Ishimori:2014jwa}.

In the present paper we propose a model which selects the following ordering,
\begin{align}
Q=
V_Q
\begin{pmatrix}
1&0&0\\
0&-1&0\\
0&0&1
\end{pmatrix}
V_Q^\dagger,
\quad
A=
V_A
\begin{pmatrix}
-1&0&0\\
0&1&0\\
0&0&1
\end{pmatrix}
V_A^\dagger,
\end{align}
where 
\begin{eqnarray}
\begin{split}
\label{qandamix}
V_Q
=\frac{1}{\sqrt2}\begin{pmatrix}
\eta^{-kx}&-\eta^{-kx}&0\\
1&1&0\\
0&0&\sqrt2
\end{pmatrix}
\begin{pmatrix}
\cos\theta&0&\sin\theta e^{i\alpha}\\
0&1&0\\
-\sin\theta e^{-i\alpha}&0&\cos\theta
\end{pmatrix},
\\
V_A
=\frac{1}{\sqrt2}
\begin{pmatrix}
-\eta^{-ly} &\eta^{-ly}&0\\
1&1&0\\
0&0&\sqrt2
\end{pmatrix}
\begin{pmatrix}
1&0&0\\
0&\cos\theta'&\sin\theta' e^{i\beta}\\
0&-\sin\theta' e^{-i\beta}&\cos\theta'
\end{pmatrix}.
\end{split}
\end{eqnarray}
For simplicity, we consider $\alpha=\beta=0$ 
in this section.
As noted above, 
the CKM matrix is given by 
$V_\text{CKM}=V_Q^\dagger V_A$ up to phase transformations so that
\begin{eqnarray}
V_\text{ CKM}=
\frac12
\begin{pmatrix}
(1-\eta^{kx-ly})c&(1+\eta^{kx-ly})cc'+2ss'&-2sc'+(1+\eta^{kx-ly})cs'\\
1+\eta^{kx-ly}&(1-\eta^{kx-ly})c'&(1-\eta^{kx-ly})s'\\
(1-\eta^{kx-ly})s&(1+\eta^{kx-ly})sc'-2cs'
&2cc'+(1+\eta^{kx-ly})ss'
\end{pmatrix},
\end{eqnarray}
where 
$c=\cos\theta$, 
$s=\sin\theta$, 
$c'=\cos\theta'$, 
and $s'=\sin\theta'$. 
If we take $N=7$,  $kx-ly=5$ with $s=-0.0021$ and $s'=0.042$,
we obtain $|V_{us}|=0.222$, $|V_{cb}|=0.0409$,
$|V_{ub}|=0.00911$ 
and $J={\rm Im}(V_{us}V_{cb}V_{ub}^*V_{cs}^*)=1.81\times 10^{-5}$ .
Detail numerical discussions will be presented  based on our model in section 5.


\section{Model building}
\subsection{Particle contents and charge assignment}
Let us present the model, which realizes the quark mass matrices with the symmetric property
 in the section \ref{3}.
As seen Table 1, we suppose the  charge assignment of the quarks and scalar fields $\chi$s 
in the flavor symmetry  $\Delta(6N^2)$ and $Z_{N+1}$ 
 where  $N/3$ is not integer.  
\begin{table}[h]
{\footnotesize
\begin{tabular}{|c|cccccccccccc|}
\hline
&$(q_1,q_2,q_3)$&$(u^c,c^c)$&$t^c$
&$(d^c,s^c)$&$b^c$ &$h_u,h_d$ 
&$\chi_{u}$&$\chi_{u}'$&$\chi_{u}''$
&$\chi_{d}$&$\chi_{d}'$&$\chi_{d}''$
\\ 
\hline
$\Delta(6N^2)$ & ${\bf 3}_{1k}$& ${\bf 2}$& ${\bf 1}_0$
& ${\bf 2}$& ${\bf 1}_0$ & ${\bf 1}_0$
&${\bf 3}_{1(-k)}$
& ${\bf 3}_{1(-k)}$& ${\bf 1}_0$& ${\bf 3}_{1(-k)}$
& ${\bf 3}_{1(-k)}$& ${\bf 1}_0$
\\
$Z_{N+1}$ & $0$& $1$& $1$
& $0$&$0$& $0$
& $-1$& $-1$& $0$& $1$& $0$& $-1$
\\
$Z_{N+1}'$ & $0$& $0$& $0$
& $1$&$1$& $0$
& $1$& $0$& $-1$& $-1$& $-1$& $0$
\\
$U(1)_R$ & $1$& $1$& $1$
& $1$&$1$& $0$
& $0$& $0$& $0$& $0$& $0$& $0$
\\
\hline
\end{tabular}
}
\caption{Particle contents and charge assignment of the flavor symmetry 
for fermions and scalar fields $\chi$'s.}
\end{table}

The superpotential for the quark sector is
\begin{eqnarray}
\begin{split}
w_q
=
&y_{u1}
((u^c+\omega^2 c^c)q_1\chi_{u1}
+(u^c+c^c)\omega q_2\chi_{u2}
+(\omega^2 u^c+c^c)q_3\chi_{u3}
)h_u\chi_u''/\Lambda^2
\\&+y_{u2}
((u^c+\omega^2 c^c)q_1\chi_{u1}'
+(u^c+c^c)\omega q_2\chi_{u2}'
+(\omega^2 u^c+c^c)q_3\chi_{u3}'
)h_u/\Lambda
\\
&+y_{u3}
t^c(q_1\chi_{u1}
+q_2\chi_{u2}
+q_3\chi_{u3}
)h_u\chi_u''/\Lambda^2
+y_{u4}
t^c(q_1\chi_{u1}'
+q_2\chi_{u2}'
+q_3\chi_{u3}'
)h_u/\Lambda
\\
&+y_{d1}
((d^c+\omega^2 s^c)q_1\chi_{d1}
+(d^c+s^c)\omega q_2\chi_{d2}
+(\omega^2 d^c+s^c)q_3\chi_{d3}
)h_d\chi_d''/\Lambda^2
\\&+y_{d2}
((d^c+\omega^2 s^c)q_1\chi_{d1}'
+(d^c+s^c)\omega q_2\chi_{d2}'
+(\omega^2 d^c+s^c)q_3\chi_{d3}'
)h_d/\Lambda
\\
&+y_{d3}
b^c(q_1\chi_{d1}
+q_2\chi_{d2}
+q_3\chi_{d3}
)h_d\chi_d''/\Lambda^2
+y_{d4}
b^c(q_1\chi_{d1}'
+q_2\chi_{d2}'
+q_3\chi_{d3}'
)h_d/\Lambda.
\end{split}
\end{eqnarray}
Multiplication rule of the group $\Delta(6N^2)$ is 
based on the review \cite{Ishimori:2010au}. 
For instance, the term of $y_{u1}$ is given by using 
$(x_1,x_2,x_3)_{3_{1k}} \times (y_1,y_2,y_3)_{3_{1k}}
=( x_1y_1+x_2y_2+x_3y_3)_{1_0}
+(\omega x_1y_1+x_2y_2+\omega^2 x_3y_3,\omega^2 x_1y_1+x_2y_2+\omega x_3y_3)_{2}
+(x_3y_2,x_1y_3,x_2y_1,x_1y_2,x_3y_1,x_2y_3)_{6_{[k,k]}}$ 
and $(x_1,x_2)_{2} \times (y_1,y_2)_{2}
=( x_1y_2+x_2y_1)_{1_0}
+( x_1y_2-x_2y_1)_{1_1}
+(x_2y_2,x_1y_1)_{2}$, where $\omega$ is the cubic 
root of one. 
The vacuum alignment is taken as
\begin{eqnarray}
\begin{split}
\langle\chi_u\rangle
=\begin{pmatrix}
u_u\\
-u_u\eta^{x}\\
0
\end{pmatrix},
\quad
\langle\chi_u'\rangle
=\begin{pmatrix}
0\\
0\\
u_u'
\end{pmatrix},
\quad
\langle\chi_u''\rangle=u_u'' ,
\\
\langle\chi_d\rangle
=\begin{pmatrix}
u_d\\
-u_d \eta^y\\
0
\end{pmatrix},
\quad
\langle\chi_d'\rangle
=\begin{pmatrix}
0\\
0\\
u_d'
\end{pmatrix},
\quad
\langle\chi_d''\rangle=u_d''.
\end{split}
\end{eqnarray}
We will discuss how to get this vacuum alignment in subsection 4.2. 
The minus signs of the vacuum expectation values (VEV's) $\langle \chi_u \rangle$
and $\langle \chi_d \rangle$ are important to get the stable vacuum in the potential 
analysis, and those can be given only when $N$ is even. 
Although, $N=7$ is the minimum number to get the 
Cabibbo angle $\theta_{12}\approx 0.22$, 
we have to take $N=14$ as the minimum to  realize the stable vacuum
in our model.
In this paper, we assume VEV's are real.
By choosing proper $Q$ and $A$, 
we can obtain $Q\langle\chi_u\rangle=\langle\chi_u\rangle$, 
$Q\langle\chi_u'\rangle=\langle\chi_u'\rangle$, 
$A\langle\chi_d\rangle=\langle\chi_d\rangle$, 
and $A\langle\chi_d'\rangle=\langle\chi_d'\rangle$ 
from the Eqs. (\ref{qanda}) when $kx=x+N/2$ and $ly=y+N/2$.
Then we have residual symmetry $Z_2\times Z_2$ 
for mass matrices of quarks. 
Actually, the mass matrices are expressed by
\begin{eqnarray}
\begin{split}
(M_u)_{RL}
&=\frac{v_u}{\Lambda^2}
\begin{pmatrix}
y_{u1}u_uu_u''&-\omega y_{u1}u_uu_u''\eta^x&\omega^2 y_{u2}u_u'\Lambda\\
\omega^2 y_{u1}u_uu_u''&-\omega y_{u1}u_uu_u''\eta^x&y_{u2}u_u'\Lambda\\
y_{u3}u_uu_u''&-y_{u3}u_uu_u'' \eta^x&y_{u4}u_u'\Lambda\\
\end{pmatrix},
\\
(M_d)_{RL}
&=\frac{v_d}{\Lambda^2}
\begin{pmatrix}
y_{d1}u_du_d''&-\omega y_{d1}u_du_d''\eta^y&\omega^2 y_{d2}u_d'\Lambda\\
\omega^2 y_{d1}u_du_d''&-\omega y_{d1}u_du_d''\eta^y&y_{d2}u_d'\Lambda\\
y_{d3}u_du_d''&-y_{d3}u_du_d''\eta^y &y_{d4}u_d'\Lambda\\
\end{pmatrix}.
\end{split}
\end{eqnarray}
They satisfy $Q^\dagger M_u^\dagger M_u Q=M_u^\dagger M_u$ 
and $A^\dagger M_d^\dagger M_d A=M_d^\dagger M_d$. 

Mass matrices in $LL$ basis become
\begin{eqnarray}
\begin{split}
&V_{12}^{u\dagger} M_u^\dagger M_u V_{12}^u
\\
&=\frac{v_u^2}{\Lambda^4}
\begin{pmatrix}
(| y_{u1}| ^2+2 |y_{u3}| ^2) u_u^2u_u''^2&0
&-\sqrt2 (y_{u1}^*y_{u2}- y_{u3}^* y_{u4})u_uu_u'u_u''\Lambda 
\\
0&3 |y_{u1} |^2u_u^2 u_u''^2&0
\\
-\sqrt2 (y_{u1}y_{u2}^*- y_{u3} y_{u4}^*)u_uu_u'u_u''\Lambda&0
&(2 |y_{u2}| ^2 + | y_{u4} |^2)u_u'^2\Lambda^2 \\
\end{pmatrix},
\\
\\
&V_{12}^{d\dagger} M_d^\dagger M_d V_{12}^d
\\
&=\frac{v_d^2}{\Lambda^4}
\begin{pmatrix}
3 |y_{d1}|^2u_d^2 u_d''^2&0&0
\\
0&(|y_{d1}|^2+2|y_{d3}|^2) u_d^2u_d''^2
&\sqrt2 (y_{d1}^*y_{d2}- y_{d3}^* y_{d4})u_du_d'u_d''\eta^{-y}
\\
0&\sqrt2 (y_{d1}y_{d2}^*- y_{d3} y_{d4}^*)u_du_d'u_d''\Lambda\eta^{y}
&(2 |y_{d2}|^2 + | y_{d4}|^2)u_d'^2\Lambda^2 \\
\end{pmatrix},
\end{split}
\end{eqnarray}
where
\begin{eqnarray}
\begin{split}
V_{12}^u
=\begin{pmatrix}
1&\eta^{x}&0\\
-\eta^{-x}&1&0\\
0&0&\sqrt2\\
\end{pmatrix}
=
\begin{pmatrix}
1&0&0\\
0&-\eta^{-kx-x}&0\\
0&0&1\\
\end{pmatrix}
V_Q
\begin{pmatrix}
\eta^{kx}&0&0\\
0&-\eta^{kx+x}&0\\
0&0&1\\
\end{pmatrix},
\\
V_{12}^{d}
=\frac{1}{\sqrt2}\begin{pmatrix}
1&-\eta^{y}&0\\
\eta^{-y}&1&0\\
0&0&\sqrt2\\
\end{pmatrix}
=\begin{pmatrix}
-1&0&0\\
0&\eta^{-ly-y}&0\\
0&0&1\\
\end{pmatrix}
V_A
\begin{pmatrix}
\eta^{ly}&0&0\\
0&\eta^{ly+y}&0\\
0&0&1\\
\end{pmatrix},
\end{split}
\end{eqnarray}
where $V_Q$ and $V_A$ are the ones of Eq. (\ref{qandamix})
(where the CKM matrix is only specified by the symmetry up to phase transformations).
Each mass matrix contains 
four parameters, we can obtain 
three masses and additional mixing angle in general.

\subsubsection{Masses and mixing}
Masses and mixing angles can be obtained by 
diagonalizing the mass matrices. 
Masses are expressed by
\begin{eqnarray}
\begin{split}
&m_u^2=\frac{v_u^2}{2\Lambda^4}
(m_{u22}^4+m_{u33}^4
-\sqrt{(m_{u22}^4-m_{u33}^4)^2
+4m_{u23}^8}~),
\quad
m_c^2=
\frac{3|y_{u1}|^2v_u^2u_u^2u_u''^2}{\Lambda^4},
\\&
m_t^2=
\frac{v_u^2}{2\Lambda^4}
(m_{u22}^4+m_{u33}^4
+\sqrt{(m_{u22}^4-m_{u33}^4)^2
+4m_{u23}^8}~),
\\
&
m_d^2=\frac{3|y_{d1}|^2v_d^2u_d^2u_d''^2}{\Lambda^4},
\quad
m_s^2=
\frac{v_d^2}{2\Lambda^4}
(m_{d22}^4+m_{d33}^4
-\sqrt{(m_{d22}^4-m_{d33}^4)^2
+4m_{d23}^8}~),
\\&
m_b^2=
\frac{v_d^2}{2\Lambda^4}
(m_{d22}^4+m_{d33}^4
+\sqrt{(m_{d22}^4-m_{d33}^4)^2
+4m_{d23}^8}~),
\end{split}
\end{eqnarray}
where
$m_{\alpha22}^4=(|y_{\alpha1}|^2+2|y_{\alpha3}|^2) u_\alpha^2u_\alpha''^2$, 
$m_{\alpha23}^4=\sqrt2 |(y_{\alpha1}^*y_{\alpha2}- y_{\alpha3}^* y_{\alpha4})|u_\alpha u_\alpha'u_\alpha''\Lambda$,  and
$m_{\alpha33}^4=(2|y_{\alpha2}|^2 + |y_{\alpha4}|^2)u_\alpha'^2\Lambda^2$
with $\alpha=u,d$. 
Similarly, mixing matrices are
\begin{eqnarray}
V^u=V_{12}^u
\begin{pmatrix}
\cos\theta_u&0&-e^{i\phi_u}\sin\theta_u \\
0&1&0\\
e^{-i\phi_u}\sin\theta_u&0&\cos\theta_u\\
\end{pmatrix},
\quad
V^d=V_{12}^d
\begin{pmatrix}
1&0&0\\
0&\cos\theta_d&-e^{i\phi_d}\sin\theta_d \\
0&e^{-i\phi_d}\sin\theta_d&\cos\theta_d\\
\end{pmatrix}.
\end{eqnarray}
where
\begin{eqnarray}
\tan2\theta_u
=\frac{2m_{u23}^4}{m_{u33}^4-m_{u22}^4},
\quad
\tan2\theta_d
=\frac{2m_{d23}^4}{m_{d33}^4-m_{d22}^4},
\end{eqnarray}
and $\phi_{u,d}$ are given by phases of Yukawa coupling 
and $\eta^y$. 
The CKM matrix is given by 
$V_\text{CKM}=V_u^\dagger V_d$ so that
\begin{eqnarray}
\begin{split}
&V_\text{ CKM}
\\
&=\frac12
\begin{pmatrix}
(1-\eta^{x-y})c_u&-(\eta^{x}+\eta^{y})c_uc_d+2e^{i(\phi_u-\phi_d)}s_us_d
&2e^{i\phi_u}s_uc_d+(\eta^x+\eta^{y})e^{i\phi_d}c_us_d\\
\eta^{-x}+\eta^{-y}
&(1-\eta^{-x+y})c_d&-(1-\eta^{-x+y})e^{i\phi_d}s_d\\
-(1-\eta^{x-y})e^{-i\phi_u}s_u
&(\eta^{x}+\eta^{y})e^{-i\phi_u}s_uc_d+2e^{-i\phi_d}c_us_d
&2c_uc_d-(\eta^{x}+\eta^{y})e^{-i(\phi_u-\phi_d)}s_us_d
\end{pmatrix},
\end{split}
\end{eqnarray}
where $s_u=\sin\theta_u$, $c_u=\cos\theta_u$, 
$s_d=\sin\theta_d$, and $c_d=\cos\theta_d$. 
For example, if we take $N=7$, $x-y=4$,
$s_u=-0.0021$ and $s_d=0.042$ with  real Yukawa couplings,
we obtain the desired values $|V_{us}|=0.22$ and $|V_{cb}|=0.041$, but
undesired one  $|V_{ub}|=0.0091$, which is
 the predicted lower bound of $|V_{ub}|$.
 
In addition, they obtain
 the unitarity triangle with three angles
$\alpha=90^\circ$, $\beta=77^\circ$, and $\gamma=13^\circ$,
which is an unfavored triangle.
 Therefore, we need to take complex Yukawa 
couplings in order to get the proper $|V_{ub}|$ and CP phase.

\subsection{Potential analysis}
\begin{table}[h]
{\small
\begin{tabular}{|c|cccccccccccccc|}
\hline
 &$\chi_{u}$&$\chi_{u}'$&$\chi_{u}''$
 &$\chi_{d}$&$\chi_{d}'$&$\chi_{d}''$
  &$\Phi_{1}$&$\Phi_{2}$&$\Phi_{3}$
 &$\Phi_{4}$&$\Phi_{5}$&$\Phi_{6}$
 &$\Phi_{7}$&$\Phi_{8}$\\ 
\hline
$\Delta(6N^2)$ 
& ${\bf 3}_{1(-k)}$& ${\bf 3}_{1(-k)}$& ${\bf 1}_0$
& ${\bf 3}_{1(-k)}$&  ${\bf 3}_{1(-k)}$& ${\bf 1}_0$
& ${\bf 1}_{0}$& ${\bf 1}_0$& ${\bf 1}_{0}$
& ${\bf 1}_{0}$& ${\bf 1}_0$& ${\bf 1}_{0}$
& ${\bf 1}_{0}$& ${\bf 1}_0$
\\
$Z_{N+1}$ 
& $-1$& $-1$&  $0$&  $1$&  $0$&  $-1$
&  $3$&  $3$&  $3$&  $-3$
&  $-1$&  $0$&  $-1$&  $1$
\\
$Z_{N+1}'$ 
& $1$& $0$&  $-1$&  $-1$&  $-1$&  $0$
&  $-3$&  $-1$&  $0$&  $3$
&  $3$&  $3$&  $1$&  $-1$
\\
$U(1)_{R}$ 
& $0$& $0$&  $0$&  $0$&  $0$&  $0$
& $2$& $2$&  $2$&  $2$
& $2$& $2$&  $2$&  $2$
\\
\hline
\end{tabular}
}
\caption{Particle contents and charge assignment of the flavor symmetry  and $U(1)_R$
for the flavon fields $\chi$'s and driving fields $\Phi_i$.}
\end{table}

In order to get desired vacuum expectation values of $\chi$'s, we introduce
 the driving fields  $\Phi_i$  with the $U(1)_R$ symmetry
in the framework of the supersymmetry.
 The charge assignment 
for the scalar fields $\chi$'s and driving fields $\Phi_i$ is given in Table 2.
Then, the leading order of the superpotential is given by
\begin{eqnarray}
\begin{split}
w
=&\frac{\lambda_1}{\Lambda}\chi_u^3\Phi_1
+\frac{\lambda_2}{\Lambda}\chi_u\chi_u'^2\Phi_2
+\frac{\lambda_3}{\Lambda}\chi_u'^3\Phi_3
+\frac{\lambda_4}{\Lambda}\chi_d^3\Phi_4
+\frac{\lambda_5}{\Lambda}\chi_d\chi_d'^2\Phi_5
+\frac{\lambda_6}{\Lambda}\chi_d'^3\Phi_6
\\
&+\sum_n
(\frac{\lambda_{7n}}{\Lambda^{2n-1}}
\chi_u^{n}\chi_d^{n+1})\Phi_7
+\frac{\lambda_{7}'}{\Lambda^{N-2}}
\chi_u^{N}\Phi_7
+\sum_n
(\frac{\lambda_{8n}}{\Lambda^{2n-1}}
\chi_u^{n+1}\chi_d^{n})\Phi_8
+\frac{\lambda_{8}'}{\Lambda^{N-2}}
\chi_d^{N}\Phi_8.
\end{split}
\end{eqnarray}
They can be explicitly written as
\begin{eqnarray}
\begin{split}
w
=&\frac{\lambda_1}{\Lambda}\chi_{u1}\chi_{u2}\chi_{u3}\Phi_1
+\frac{\lambda_2}{\Lambda}
(\chi_{u1}\chi_{u2}'\chi_{u3}'
+\chi_{u2}\chi_{u1}'\chi_{u3}'
+\chi_{u3}\chi_{u1}'\chi_{u2}')\Phi_2
+\frac{\lambda_3}{\Lambda}
\chi_{u1}'\chi_{u2}'\chi_{u3}'\Phi_3
\\
&+\frac{\lambda_4}{\Lambda}\chi_{d1}\chi_{d2}\chi_{d3}\Phi_4
+\frac{\lambda_5}{\Lambda}
(\chi_{d1}\chi_{d2}'\chi_{d3}'
+\chi_{d2}\chi_{d1}'\chi_{d3}'
+\chi_{d3}\chi_{d1}'\chi_{d2}')\Phi_5
+\frac{\lambda_6}{\Lambda}
\chi_{d1}'\chi_{d2}'\chi_{d3}'\Phi_6
\\
&+\sum_{n_1,n_2,n_1\geq n_2}
\frac{\lambda_{7n_1,n_2}}{\Lambda^{12n_1-6n_2+1}}
(\chi_{u1}\chi_{u2}\chi_{u3})^{n_1}
(\chi_{d1}\chi_{d2}\chi_{d3})^{n_2}
\\
&\qquad
\times (\chi_{u1}\chi_{d2}\chi_{d3}
+\chi_{u2}\chi_{d1}\chi_{d3}
+\chi_{u3}\chi_{d1}\chi_{d2})^{3n_1-3n_2+1}
\Phi_7
\\
&+\sum_{n_1,n_2,n_1\geq n_2}
\frac{\lambda_{8n_1,n_2}}{\Lambda^{12n_1-6n_2+1}}
(\chi_{u1}\chi_{u2}\chi_{u3})^{n_1}
(\chi_{d1}\chi_{d2}\chi_{d3})^{n_2}
\\
&\qquad
\times(\chi_{u1}\chi_{u2}\chi_{d3}
+\chi_{u2}\chi_{u3}\chi_{d1}
+\chi_{u3}\chi_{u1}\chi_{d2})^{3n_1-3n_2+1}
\Phi_8
\\
&+\frac{\lambda_7'}{\Lambda^{N-2}}
(\chi_{u1}^N+\chi_{u2}^N+\chi_{u3}^N)\Phi_7
+\frac{\lambda_8'}{\Lambda^{N-2}}
(\chi_{d1}^N+\chi_{d2}^N+\chi_{d3}^N)\Phi_8.
\end{split}
\end{eqnarray}
By solving the potential minimum conditions, we obtain
the vacuum expectation values as follows:
\begin{eqnarray}
\begin{split}
\langle\chi_u\rangle
=\begin{pmatrix}
u_u\\
-u_u\eta^{x}\\
0
\end{pmatrix},
\quad
\langle\chi_u'\rangle
=\begin{pmatrix}
0\\
0\\
u_u'
\end{pmatrix},
\quad
\langle\chi_d\rangle
=\begin{pmatrix}
u_d\\
-u_d \eta^y\\
0
\end{pmatrix},
\quad
\langle\chi_d'\rangle
=\begin{pmatrix}
0\\
0\\
u_d'
\end{pmatrix},
\end{split}
\end{eqnarray}
where  $N$ is taken to be  even otherwise 
the minus sign does not appear.
These VEV's present desirable vacuum alignments.

\subsection{$Z_2$ breaking terms}
$Z_2$ breaking terms for the Yukawa couplings are 
highly suppressed. The leading order for the breaking is
\begin{eqnarray}
\begin{split}
\Delta w_q
=&y_{b1}((u^c+\omega^2 c^c)q_1\chi_{d1}
+(u^c+c^c)\omega q_2\chi_{d2}
+(\omega^2 u^c+c^c)q_3\chi_{d3}
)h_u\chi_u''^{N-1}\chi_d''^2/\Lambda^{N+2}
\\
&+y_{b2}
t^c(q_1\chi_{d1}
+q_2\chi_{d2}
+q_3\chi_{d3}
)h_u\chi_u''^{N-1}\chi_d''^2/\Lambda^{N+2}
\\
&+y_{b3}
((d^c+\omega^2 s^c)q_1\chi_{u1}
+(d^c+s^c)\omega q_2\chi_{u2}
+(\omega^2 d^c+s^c)q_3\chi_{u3}
)h_d\chi_u''^2\chi_d''^{N-1}/\Lambda^{N+2}
\\
&+y_{b4}
b^c(q_1\chi_{u1}
+q_2\chi_{u2}
+q_3\chi_{u3}
)h_d\chi_u''^2\chi_d''^{N-1}/\Lambda^{N+2}.
\end{split}
\end{eqnarray}
For the superpotential of scalar fields, the leading order of 
$Z_2$ breaking terms appears as
\begin{eqnarray}
\begin{split}
\Delta w
=&\frac{\lambda_{b1}}{\Lambda^{N-1}}\chi_u'^N\chi_u''\Phi_7
+\frac{\lambda_{b2}}{\Lambda^{N-1}}\chi_d'^N\chi_d''\Phi_8.
\end{split}
\end{eqnarray}
The VEV's of $\chi_u$ and $\chi_d$  are deviated by these terms.
Then, the vacuum alignment is deviated by
\begin{eqnarray}
\begin{split}
\langle\chi_u\rangle
=\begin{pmatrix}
u_u+{\cal O}({u_u'^{N}u_u''}/{\Lambda^{N}})\\
-u_u\eta^{x}+{\cal O}({u_u'^{N}u_u''}/{\Lambda^{N}})\\
0
\end{pmatrix},
\quad
\langle\chi_d\rangle
=\begin{pmatrix}
u_d+{\cal O}({u_d'^{N}u_d''}/{\Lambda^{N}})\\
-u_d \eta^y+{\cal O}({u_d'^{N}u_d''}/{\Lambda^{N}})\\
0
\end{pmatrix},
\end{split}
\end{eqnarray}
and alignment of other fields are highly suppressed. 
With this deviation, the mass matrix is modified as
\begin{eqnarray}
\label{Z2break}
\begin{split}
(M_u)_{RL}
&=\frac{v_u}{\Lambda^2}
\begin{pmatrix}
y_{u1}u_uu_u''&-\omega y_{u1}u_uu_u''\eta^x&\omega^2 y_{u2}u_u'\Lambda\\
\omega^2 y_{u1}u_uu_u''&-\omega y_{u1}u_uu_u''\eta^x&y_{u2}u_u'\Lambda\\
y_{u3}u_uu_u''&-y_{u3}u_uu_u'' \eta^x&y_{u4}u_u'\Lambda\\
\end{pmatrix}
+\frac{v_u}{\Lambda^{N+2}}
\begin{pmatrix}
{\cal O}(u_u'^Nu_u''^2)&{\cal O}(u_u'^Nu_u''^2)&0\\
{\cal O}(u_u'^Nu_u''^2)&{\cal O}(u_u'^Nu_u''^2)&0\\
{\cal O}(u_u'^Nu_u''^2)&{\cal O}(u_u'^Nu_u''^2)&0\\
\end{pmatrix},
\\
(M_d)_{RL}
&=\frac{v_d}{\Lambda^2}
\begin{pmatrix}
y_{d1}u_du_d''&-\omega y_{d1}u_du_d''\eta^y&\omega^2 y_{d2}u_d'\Lambda\\
\omega^2 y_{d1}u_du_d''&-\omega y_{d1}u_du_d''\eta^y&y_{d2}u_d'\Lambda\\
y_{d3}u_du_d''&-y_{d3}u_du_d''\eta^y &y_{d4}u_d'\Lambda\\
\end{pmatrix}
+\frac{v_d}{\Lambda^{N+2}}
\begin{pmatrix}
{\cal O}(u_d'^Nu_d''^2)&{\cal O}(u_d'^Nu_d''^2)&0\\
{\cal O}(u_d'^Nu_d''^2)&{\cal O}(u_d'^Nu_d''^2)&0\\
{\cal O}(u_d'^Nu_d''^2)&{\cal O}(u_d'^Nu_d''^2)&0\\
\end{pmatrix}.
\end{split}
\end{eqnarray}
Thus,  the magnitude of $Z_2$ breaking terms 
for the mass matrix is of order ${\cal O}(u_u'^Nu_u''/u_u\Lambda^N)$ 
for up-type quarks and ${\cal O}(u_d'^Nu_d''/u_d\Lambda^N)$ for down-quarks,
respectively.

\section{Numerical analysis}
When the subgroup $Z_2\times Z_2$ is preserved and 
the phase of VEV's is fixed, the number of parameters 
is four in each mass matrix. 
Then we can obtain three masses and one mixing angle 
as free parameters. 
For the symmetry and phases, we choose $N=14$ and 
$x-y=6$ then we predict 
$\sin\theta_{12}=0.222521$ at the leading order. 
\footnote{This is identical to the example in the Introduction for $N=7$ as well as 
the prediction of the previous model for $N=28$ \cite{Ishimori:2014jwa}.}
This is to be compared to the experimental value at the weak scale of 
$|V_{us}|=0.225\pm 0.001$.

Suppose that the flavor symmetry exists at the scale of the  grand unified theory (GUT).
Then, we should fit the quark masses and mixing angles at the GUT scale with the supersymmetry. 
Inputting experimental data at the  low energy scale,
the  renormalization group runnings give us
following values  \cite{Antusch:2013jca}:
\begin{eqnarray}
\begin{split}
\label{eqreg}
\theta_{12}\approx 0.2276,
\quad
2.9\times 10^{-3}\leq\theta_{13}\leq
3.4\times 10^{-3},
\quad
3.3\times 10^{-2}\leq\theta_{23}\leq
3.9\times 10^{-2},
\\
4.8\times 10^{-6}\leq\frac{m_u}{m_t}
\leq 5.4\times 10^{-6},
\quad
2.3\times 10^{-3}\leq\frac{m_c}{m_t}
\leq2.6\times 10^{-3},
\\
6.3\times10^{-4}\leq\frac{m_d}{m_b}
\leq 8.9\times10^{-4},
\quad
1.8\times10^{-2}\leq\frac{m_s}{m_b}
\leq 1.2\times10^{-2}.
\end{split}
\end{eqnarray}
We reproduce these mass and mixing angles by
 scattering our model parameters while $N=14$ and $x-y=6$ are fixed.

In the Figures 1 and 2,
we show the scattering plots 
to see the consistency with experiments. 
Giving random values for all the Yukawa couplings with phases and
VEV's of flavons, we get quark masses and mixing angles by 
diagonalising mass matrices of up- and down-type quarks, which are 
 constrained by the observed  values in Eq. (\ref{eqreg}). 
The physical values are actually three up-quark masses, three-down quark masses, three mixing angles, and CP phase. 
Since the third generation masses can be determined independently, we fit the  mass ratios.

For the case of  the $Z_2\times Z_2$ invariant quark mass matrices,  we plot
the  CKM matrix elements, the CP angles ($\alpha$, $\beta$, $\gamma$)
and the mass ratios  in Figure 1,
where  red and blue cross marks denote the experimental central values
at the weak scale \cite{pdg} since the running effect is small.

As discussed  above,  $|V_{us}|$ ($|V_{cd}|$) is predicted  to be in the very narrow range
even if the next leading terms are added to the leading term $|\eta^{x}+\eta^{y}|/2$.
The  CKM elements $|V_{cb}|$  and $|V_{ts}|$  are reproduced due to  the parameter $\theta_d$.
 The $|V_{ub}|$ and $|V_{td}|$  depend on  both $\theta_d$  and $\theta_u$.
 Due to the phases of Yukawa couplings, these  elements are fitted well.
The three angles of the unitarity triangle and the quark mass ratios are also  reproduced.

\begin{figure}
\includegraphics[width=8cm]{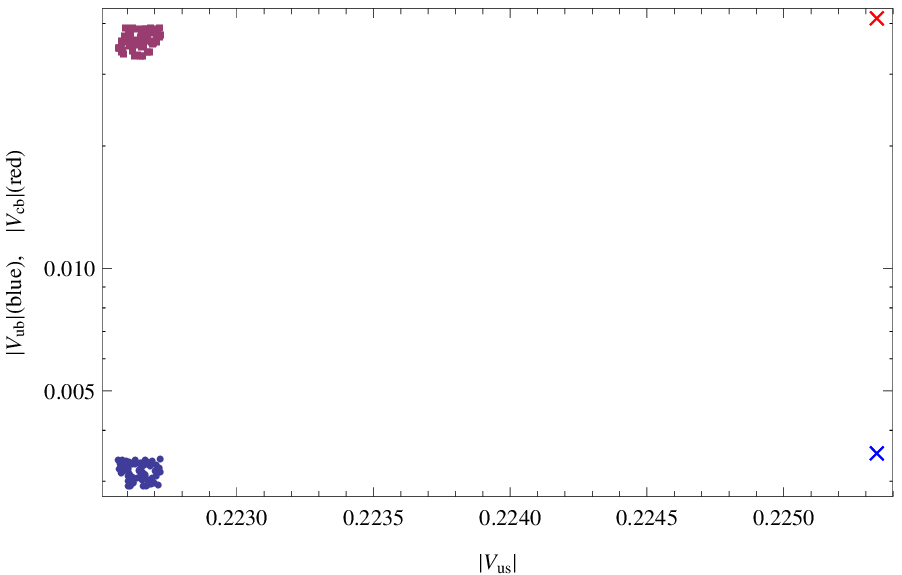}
\quad
\includegraphics[width=8cm]{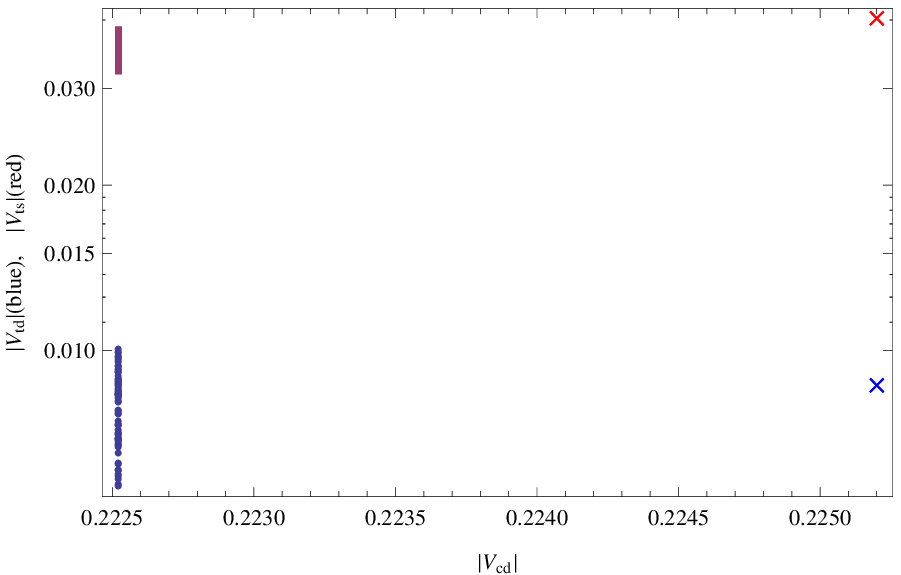}
\\
\includegraphics[width=8cm]{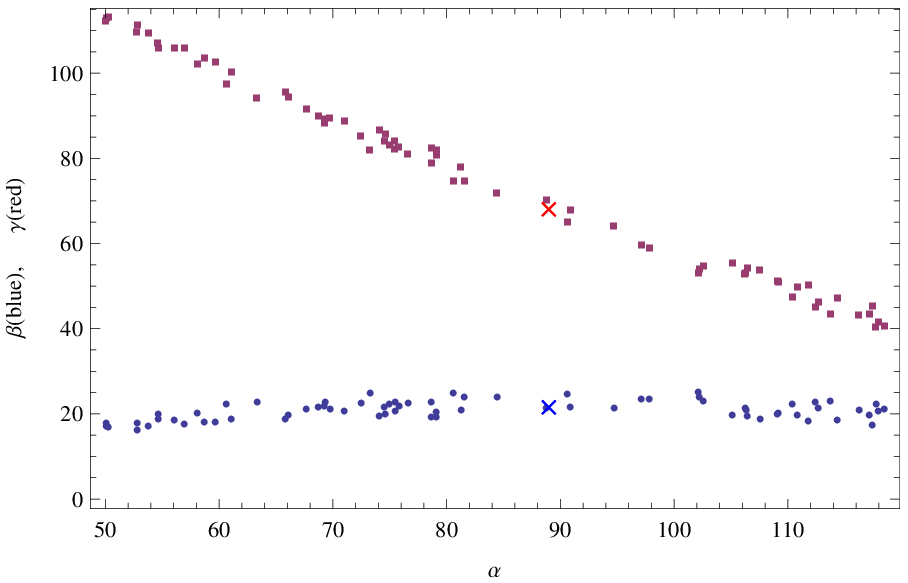}
\quad
\includegraphics[width=8cm]{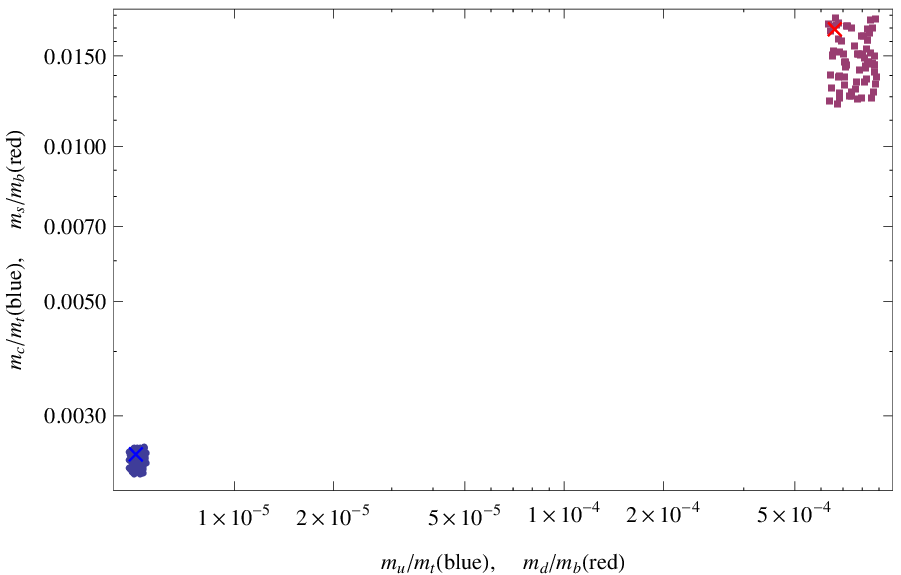}
\caption{Scattering plots among the CKM matrix elements,   
 the angles of the unitarity triangle and the mass ratios
 in the case of the $Z_2\times Z_2$ invariant mass matrices.
Cross marks denote the experimental central values.}
\end{figure}


In order to fit the mixing angles perfectly, especially  $|V_{us}|$,
$Z_2$ breaking terms are required.
As seen in  Eq. (\ref{Z2break}),  the mass matrices 
are  modified due to the deviation of VEV's. 
Comparing to the leading terms that preserve $Z_2$, 
the magnitude of breaking terms is of order 
$u_u'^Nu_u''/u_u\Lambda^N$. 
We show the scattering plot of the CKM matrix elements 
including the  $Z_2$  breaking  effect at   3\% level
in Figure 2, where  red and blue cross marks also  denote the experimental central values
at the weak scale \cite{pdg}.
As seen in this figure, 
we can reproduce the experimental values of the mixing angles perfectly
 if $Z_2$ is broken of order $3\%$.

\begin{figure}
\includegraphics[width=8cm]{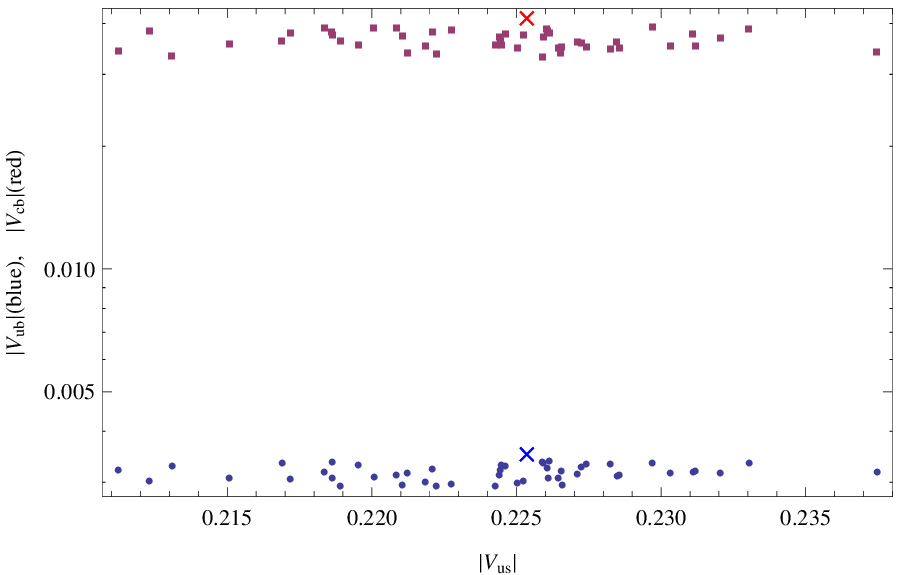}
\quad
\includegraphics[width=8cm]{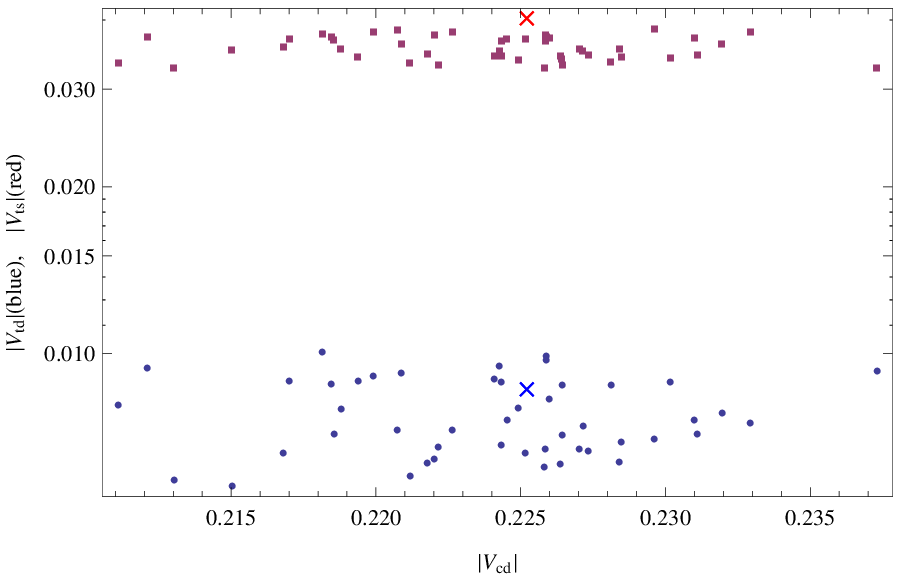}
\\
\caption{Relations among the CKM matrix elements, where
the  $Z_2$ breaking effects  of order 3\% are introduced.
Cross marks denote the experimental central values.}
\end{figure}

\section{Summary}
\label{6}
We have considered a direct approach to quark mixing based on the discrete
family symmetry $\Delta (6N^2)$
in which the Cabibbo angle is determined
by a residual $Z_2\times Z_2$ subgroup to be $|V_{us}|=0.222521$,
for $N$ being a multiple of 7. 
This prediction is very close to the experimental value 
$|V_{us}|=0.225\pm 0.001$.
We have proposed a particular model in which 
 $|V_{cb}|$,  $|V_{ub}|$ and the CP phase may occur without breaking the residual $Z_2\times Z_2$ symmetry. 
We performed a numerical analysis of the model for $N=14$, which realizes
the stable vacuum.
For the $Z_2\times Z_2$  invariant quark mass matrices,  
the  CKM matrix elements, the CP angles ($\alpha$, $\beta$, $\gamma$)
and the mass ratios  are accommodated to the experimental data.
The small $Z_2\times Z_2$ breaking effects of order 3\% 
 allow perfect agreement within the uncertainties of the 
experimentally determined best fit quark mixing values.

Finally, it is tempting to speculate that $\Delta (6N^2)$ could be suitable as a candidate family symmetry for a complete model of quark and lepton masses and mixing. In the lepton sector, $\Delta (6N^2)$ has been shown to be the only viable candidate group which can provide a direct symmetry explanation of the lepton mixing,  with a preserved Klein symmetry $Z_2\times Z_2$ in the neutrino sector and a $Z_3$ in the charged lepton sector, where both symmetries are subgroups of $\Delta (6N^2)$. However no detailed model of leptons has been proposed. Here we have proposed a $\Delta (6N^2)$ model of quarks where a different $Z_2\times Z_2$ subgroup controls the quark sector, providing an explanation of the Cabibbo angle for $N$ being a multiple of 7, while allowing a good fit to other quark mixing parameters. It might be possible to extend this model to include also leptons, although we leave this idea for future work. 

\vskip 5 cm

\section*{Acknowledgement}

This work was supported in part by the Grant-in-Aid for Scientific Research 
No.24654062 (M.T). SFK acknowledges support from the 
European Union FP7 ITN-INVISIBLES (Marie Curie Actions, PITN- GA-2011- 289442) and the STFC Consolidated ST/J000396/1 grant.

\end{document}